\documentclass[superscriptaddress,twocolumn,10pt]{revtex4-1}

\usepackage{amsmath,graphicx,color}
\usepackage{booktabs,array,dcolumn}
\usepackage{multirow}
\usepackage{times}
\usepackage{xcolor,soul}
\newcolumntype{C}{>{\centering}p}

\begin{document}

%\title{Comparing the Phase Diagram of Lead with Empirical and a Machine Learnt Interatomic Potential}

\title{Benchmarking empirical and machine-learned interatomic potentials using phase diagram predictions for Lead}

\author{Tom Hellyar}
\affiliation{Department of Chemistry, University of Warwick, Coventry, CV4 7AL, UK}

\author{Pascal T. Salzbrenner}
\affiliation{Department of Materials Science and Metallurgy, University of Cambridge, Cambridge, UK}

\author{Peter I. C. Cooke}
\affiliation{Department of Materials Science and Metallurgy, University of Cambridge, Cambridge, UK}

\author{Chris J. Pickard}
\affiliation{Department of Materials Science and Metallurgy, University of Cambridge, Cambridge, UK}
\affiliation{Advanced Institute for Materials Research, Tohoku University, Sendai, Japan}

\author{Scott Habershon}
\affiliation{Department of Chemistry, University of Warwick, Coventry, CV4 7AL, UK}

\author{Livia B. P\'artay}
\email{Livia.Bartok-Partay@warwick.ac.uk}
\affiliation{Department of Chemistry, University of Warwick, Coventry, CV4 7AL, UK}

\date{\today}

\begin{abstract}
\noindent We compare the predicted phase behaviour of lead (Pb) using three different interatomic potential models, including an embedded atom method (EAM), a modified embedded atom method (MEAM), and a neural network-based machine-learned model in the form of an ephemeral data-derived potential (EDDP). Using nested sampling and replica-exchange nested sampling simulations, we computed thermodynamic and structural properties at pressures up to 60 GPa, mapping both melting behaviour and solid-phase stability. Both the EAM and MEAM models predict the face-centred cubic (FCC) phase to remain stable up to approximately 60 GPa. In contrast, the EDDP model captures the experimentally-observed FCC-to-hexagonal close-packed (HCP) transition at around 15 GPa. These results highlight the importance of training data and model flexibility in accurately describing high-pressure phase behaviour, and demonstrate the effectiveness of nested sampling as a robust framework for exploring phase stability in materials. Particularly, the combination of nested sampling with modern machine-learned interatomic potentials -- delivering near \textit{ab initio} accuracy at tractable cost -- opens the door to truly predictive and exhaustive exploration. EDDPs trained on diverse, out-of-equilibrium configurations appear particularly well suited to this task, offering a robust and transferable framework for unbiased phase discovery.

\end{abstract}

    \maketitle

%%%%%%%%%%%%%%%%%%%%%%%%%%%%%%%%%%%%%%%%%%%%%
\section{Introduction}

Atomistic simulations play a central role in materials science, offering insights into the structural, thermodynamic, and dynamic properties of materials across a wide range of conditions. 
At the heart of these simulations lies the description of interatomic interactions, which ultimately determines both the accuracy and computational cost.
While widely-used empirical interatomic potentials -- usually reliant on predefined functional forms with relatively few fitting parameters -- offer  computational efficiency, their applicability is often limited to narrow regions of configuration space, and they may fail to reproduce even qualitative features of macroscopic behaviour when applied beyond their original training conditions~\cite{hobza1997performance,hobza1997performance,li2013effect,NS_lithium,ns_iron_2018}.
The recent emergence of machine-learned interatomic potentials (MLIPs) has dramatically softened this trade-off~\cite{behler_atom-centered_2011, rupp_fast_2012, bartok_representing_2013,behler2016perspective, deringer2019machine, bartok2017machine}. By leveraging large datasets of quantum mechanical reference calculations, MLIPs nearly match \textit{ab initio} methods' accuracy while retaining orders-of-magnitude computational speed-up over them. 
This opens up the possibility of simulating materials over longer times and larger length-scales, while capturing features that govern subtle properties and behaviour.

Validating interatomic potentials, especially MLIPs, is a fundamental step of the development process. Although a wide range of benchmarks and tools are being developed to compare structural and dynamical properties~\cite{chiang2025mlip,choudhary2024jarvis,fu2022forces}, evaluating thermodynamic behaviour -- particularly in relation to phase stability and phase transitions -- often remains neglected. This is due to both the computational cost and the complexity of accessing free energies and sampling the configuration space at finite temperatures, where entropy plays a critical role. As a result, robust thermodynamic validation is still relatively rare, despite its importance for predicting realistic material behaviour under varying pressure and temperature conditions \cite{unglert2023neural, fletcher2025autonomous}.

In this work, we apply a robust and unbiased sampling approach -- nested sampling (NS)~\cite{NS_all_review,NS_mat_review} -- to assess the phase behaviour predicted by different interatomic potentials for elemental lead. NS has proven to be a powerful tool for mapping out pressure–temperature phase diagrams~\cite{pt_phase_dias_ns,ConPresNS} because it gives direct access to the partition function and thermodynamic observables across a wide range of conditions, without prior knowledge of phases~\cite{1st_NS_paper}. NS has been successfully employed to evaluate phase stability in numerous systems described by classical empirical models~\cite{jagla,CuAu_Pastewka,ns_iron_2018,psf2022005005,NS_lithium}, 
as well as machine-learned models~\cite{AgPd_ML,ns_Pt,ns_carbon,unglert2023neural, fletcher2025autonomous}. 
We also employ the recently proposed replica-exchange nested sampling (RENS) \cite{unglert_replica_2025}, to take advantage of its enhanced ability to tackle solid-solid phase transitions. 
We use both NS and RENS to compare the predictive performance of three different potentials describing lead: (i) an established Embedded Atom Model (EAM) developed by Wang {\it et al.}~\cite{wang_improved_2018}, (ii) a Modified EAM (MEAM) published by Lee {\it et al.}~\cite{lee_semiempirical_2003}, and (iii) an MLIP based on neural networks, specifically an ephemeral data-derived potential (EDDP)~\cite{pickard2022ephemeral,salzbrenner2023} (referred to here as Pb-EDDP) which was trained on data for small unit cells and is proposed to offer good transferability across various phases across a wide range of thermodynamic conditions. 

Lead is a technologically and scientifically important heavy metal, notable for its high density, low melting point, and rich nuclear chemistry. Its isotopes frequently appear as end products of radioactive decay chains, making it relevant across disciplines ranging from nuclear science to geochemistry. 
Structurally, lead is known to adopt a face-centred cubic (FCC) phase under ambient conditions~\cite{mcmahon_high-pressure_2006,mao_high-pressure_1990}. Experimental and computational studies have shown that with increasing pressure, lead undergoes a sequence of phase transitions -- first to a hexagonal close-packed (HCP) phase around $13-14$~GPa ~\cite{mcmahon_high-pressure_2006,mao_high-pressure_1990}, followed by a transition to a body-centred cubic (BCC) phase at higher pressures (reported to be between $74$~GPa and $111$~GPa, with the free energy difference between the two phases suspected to be very small)~\cite{mcmahon_high-pressure_2006, mao_high-pressure_1990}.
This relatively simple and well-characterised phase behaviour makes lead an ideal test case for the development and benchmarking of macroscopic properties of interatomic potentials.

The remainder of this paper is organised as follows. Section~\ref{sct:methods} provides a brief overview of the NS methodology, a summary of the three interatomic models employed in this study and computational details.  Section~\ref{sct:results} highlights the similarities and differences in the predicted thermodynamic properties, discusses how high-temperature solid structures are identified and uses this information to characterise properties of the solid phase. Finally, the pressure-temperature phase diagram is presented for all three models. 
We conclude by discussing the implications for future interaction model development and validation.

%%%%%%%%%%%%%%%%%%%%%%%%%%%%%%%%%%%%%%%%%%%%%
%%%%%%%%%%%%%%%%%%%%%%%%%%%%%%%%%%%%%%%%%%%%%
%%%%%   Methods                
\section{Methods} \label{sct:methods}

%%%%%%%% Nested sampling %%%%%%%%%%%%%
\subsection{Nested sampling} \label{sct:ns_method}

NS is a Bayesian statistics technique~\cite{bib:skilling2,NS_all_review} designed to sample the probability density in high-dimensional spaces, adapted to exhaustively sample the phase space of atomistic systems~\cite{1st_NS_paper,NS_mat_review}. The unique strength of NS is that it gives direct access to the partition function, $Z$, of an $N$ particle system at pressure $p$, as a discrete sum:
\begin{equation}
  Z(N,\beta, p) = \int e^{-\beta H(\textbf{x},\textbf{p})} d\textbf{x}~d\textbf{p} \approx \sum_i(\Gamma_i - \Gamma_{i-1})e^{-\beta H(\textbf{x}_i,\textbf{p}_i)},
  \label{Z_int_micro_eq}
\end{equation}
where $\beta$ is the thermodynamic temperature, and $H$ is the enthalpy of the $i$-th sample configuration dependent on spatial and momentum coordinates, \textbf{x} and \textbf{p} respectively, although the momentum-dependent contribution can be separated out.
The heart of the NS method is an iterative algorithm that allows estimation of the phase space volume $\Gamma_i$ that bounds all the microstates with enthalpy lower than $H_i$. The algorithm has been described in detail previously~\cite{pt_phase_dias_ns,ConPresNS,NS_mat_review}; here, we provide only a brief overview.

The sampling starts by generating $K$ random configurations, evenly distributed through the entire phase space (these are usually referred to as \emph{walkers} or the \emph{live set}). During the iterative step:
\begin{enumerate}
    \item Select the highest enthalpy walker and record its enthalpy as the upper-bound limit for the current $i$-th iteration, $H_i$. 
    \item Remove this configuration from the live set.
    \item Generate a new, uniformly-random configuration, with the requirement that its enthalpy is less than $H_i$. The estimate of the phase space volume of all the accessible microstates is $\Gamma_i=\Gamma_{i-1} K / (K+1)$.
    \item Repeat, starting from step 1.
\end{enumerate}
Thermodynamic properties, such as enthalpy, $H$, and heat capacity, $C_p$, can be derived from the partition function after the sampling~\cite{1st_NS_paper,pt_phase_dias_ns}, as
\begin{equation}
H = -\frac{\partial \ln Z(N, \beta, p)}{\partial \beta} ~\mathrm{~and~}~ 
C_p=\Big(\frac{\partial H}{\partial T}\Big)_p, 
\end{equation}
and the expected value of any observable, $O$,  can be evaluated via the following relation at any arbitrary temperature:
\begin{equation}
  O(\beta) = \frac{\sum_iO_i (\Gamma_i - \Gamma_{i-1})e^{-\beta H(\textbf{x}_i,\textbf{p}_i)}}{Z(N, \beta, p)}.
\end{equation}
In the current work, we also take advantage of an extension of NS, the replica-exchange nested sampling (RENS) method, developed by Unglert {\it et al.}~\cite{unglert_replica_2025}. 
In RENS, the Hamiltonian replica-exchange concept is used to facilitate more efficient exploration of regions of the phase space separated by high energy barriers. This is achieved by allowing exchange of walker configurations between a series of NS simulations at different pressures executed in parallel. Any number of parallel NS runs can be performed within RENS, with swaps typically considered between adjacent runs, and accepted if both configurations lie below the enthalpy bound of the other sampling run. This procedure can greatly improve convergence and result in smoother predictions, particularly in challenging cases of solid-solid phase transitions, where individual NS simulations can sometimes struggle to explore both potential energy basins sufficiently. In the current work, we compare phase diagrams evaluated with NS and RENS, also providing us with useful technical insight into their performance.

%%%%%%%%%%%%%%% models %%%%%%%%
\subsection{Pb-EDDP model}

EDDPs are a recently developed neural network-based MLIP. They are referred to as `ephemeral' to emphasise that they are not intended to describe all possible chemistries of a system, let alone all elements of the periodic table as universal MLIPs aim to do~\cite{yang2024mattersim, neumann2024orb, fu2025esen, batatia2025mace}; rather, they are custom-designed for a given task, and cheap in both DFT and neural network cost~\cite{pickard2022ephemeral, salzbrenner2023}. For a typical unary system, as in this case, $\sim 6000$ DFT data points and a feed-forward neural network with a single five-node hidden layer are sufficient to attain meV-level accuracy compared to DFT.

EDDPs are trained using an approach similar to \textit{ab initio} random structure searching (AIRSS)~\cite{pickard2006silane, pickard2011airss}. The AIRSS procedure involves starting from randomly generated configurations (subject to reasonable physical constraints, such as minimum interatomic distances) and performing a geometry optimisation to find the nearest local energy minimum. This unveils a variety of minima distributed throughout the PES, in a largely uniform manner.

EDDPs use structures generated by AIRSS to iteratively build the MLIP training dataset. Importantly, the first iteration is trained on unrelaxed random structures; that is, the first frame of a typical AIRSS run. This both bootstraps the potential without any need for costly DFT geometry optimisations, and provides it with broad coverage of unfavourable high-energy structures, which are not often included in MLIP datasets. The first EDDP iteration is then used to carry out an AIRSS search. The discovered structures, alongside `shaken' versions (where atoms are slightly displaced off their equilibrium positions), are then added to the dataset. This step is typically repeated $\sim 5$ times. The geometry optimisations for the Pb-EDDP were carried out at randomly selected pressures between $0$ and $50$ GPa, and the potential was shown to describe the crystal structures and their dynamics accurately within this range~\cite{salzbrenner2023}.

Notably, EDDPs are trained on small unit cells ($2-6$ atoms in the case of the Pb-EDDP), acting as building blocks, using which the EDDP can accurately describe much larger configurations~\cite{salzbrenner2023}. As a result, DFT expense can be further minimised. This made it possible to train the Pb-EDDP on DFT energies including the relativistic spin orbit coupling, which becomes important for heavy elements~\cite{strange1998relativistic}.

It is worth remarking that AIRSS and NS share a close methodological affinity in their emphasis on the exploration of wide regions of phase space. By necessity, this involves highly non-equilibrium structures. Their inclusion in the EDDP training set, therefore, suggests EDDPs might be a particularly appropriate choice as an MLIP `backend' for NS.

\subsection{Pb-EAM and Pb-MEAM models}

We selected two classical interatomic potential models based on the EAM formalism to compare their phase properties -- the EAM developed by Wang {\it et al.}~\cite{wang_improved_2018} (selecting the more accurate EAM-II from their work) and the MEAM published by Lee {\it et al.}~\cite{lee_semiempirical_2003}. We will refer to these models as Pb-EAM and Pb-MEAM, respectively, throughout the paper. 

The Pb-EAM~\cite{wang_improved_2018} was specifically developed to study the high pressure behaviour of lead, using the lattice constant, cohesive energy, unrelaxed single vacancy formation energy, elastic constants and body moduli as well as the equation of state from reference DFT calculations. The model was found to reproduce both the FCC to HCP and HCP to BCC transitions reported, at pressures of $13.3$~GPa and $79$~GPa respectively. The equation of states demonstrates good agreement with \textit{ab initio} calculations up to $100$~GPa. The melting point of $600$~K at zero pressure aligns well with experimental data (between $600.5$~K and $601$~K)~\cite{errandonea2010melting}. We used the model in the form as it is published in the NIST Interatomic Potentials Repository~\cite{PotentialDatabase}, in the {\tt LAMMPS}~\cite{LAMMPS} potential file format.

The Pb-MEAM~\cite{lee_semiempirical_2003} is a second nearest-neighbour (2NN) potential, an improvement upon first nearest-neighbour (1NN) MEAMs, due to being able to account for second-neighbour shell interactions, which are particularly important for accurately modelling BCC systems. Previous MEAM models often resulted in FCC thermal expansion coefficients an order of magnitude lower than experimental data in the range of $273$~K to $373$~K. However, it was found that these problems could be fixed via re-parameterisation, involving careful consideration of the degree of many-body screening. Pb-MEAM was parameterised with emphasis on accurate description of elastic constants, vacancy formation energy, surface energy, FCC to HCP and FCC to BCC formation energies, and -- unlike previous MEAM models -- stacking fault energy, thermal expansion coefficients and activation energy of vacancy diffusion. However, the reported melting temperature of $705$~K under standard conditions is less accurate than the Pb-EAM model. Phase transition pressures are not explicitly described in the original work, implying that the focus was on accurate description of the FCC region and its melting line. Any further phase transitions observed at higher pressures may not be intentional. As with Pb-EAM, we obtained the potential file from the NIST Interatomic Potentials Repository~\cite{PotentialDatabase}, in the {\tt LAMMPS} file format.

%%%%%%%%%%%%%%%%%%%%%%%%%%%%%%%%%%%%%%%%%%%%%
%%%%%%%%%%%%%%%%%%%%%%%%%%%%%%%%%%%%%%%%%%%%%
%%%%%   Computational details   
\subsection{Computational details}

NS calculations were performed as presented in~\cite{pt_phase_dias_ns}, using the parallel implementation of the NS algorithm in the {\tt pymatnest} Python software
package~\cite{pymatnest}, with the {\tt LAMMPS} package~\cite{LAMMPS} (modified to include the EDDP potential~\cite{EDDPPackage}) used for the dynamics.

NS calculations were initiated using the parameters shown in Table~\ref{ns_params_table}. Although the sampling is performed using small system sizes ($N$, the number of atoms), these still allow the systematic exploration of the configuration space, capturing thermodynamically relevant phases across a wide pressure range~\cite{NS_mat_review}.
Initial configurations were generated by placing particles randomly in the simulation box. However, to avoid atoms at non-physical, short distances, where the potential is not well-behaved, an exclusion distance of $3.0$~\AA~was used in the case of the Pb-EDDP model. 
The number of walkers, $K$, controls the resolution of the sampling, with the computational cost depending linearly on this parameter. We chose $K$ such that the resulting heat capacity peaks are sufficiently converged to locate phase transitions.
At each iteration, a new walker is generated via a random walk of length $L$, performed on a randomly selected and cloned walker configuration. These random walks include changes to the volume and shape (stretch and shear steps) of the simulation cell, as well as changes to the atomic positions via short Galilean Monte Carlo trajectories~\cite{skilling_GMC,ConPresNS}, with the relative ratio of each type of move fixed at $4$:$2$:$2$:$8$, respectively. To avoid non-physical thin cells distorting the sampling, the simulation cell was not allowed to shear or stretch such that the shortest dimension becomes smaller than $80$~\% of a cubic cell of the same volume.\cite{pt_phase_dias_ns, NS_mat_review}
All NS simulations were performed until configurations required to represent properties at a temperature of 100~K were sampled sufficiently.

RENS calculations were performed in the same way as individual NS runs, albeit with careful choice of the pressure grid because the walker swap acceptance ratio depends on the shared phase space of consecutive replicas. The choice of pressure ranges for RENS was informed by the results of individual NS calculations, where clarification of solid structures was needed.

The FCC to HCP transition should be better modelled using RENS, with pressures closest to the transition line able to exchange FCC and HCP configurations more readily than those farther away. Further assisting the smoothening of transitions, a reduced spacing between pressures was used for RENS around the solid-solid transition regions to increase swapping probabilities.

\begin{table*}
\begin{center}
\caption{Parameters used for the nested sampling calculations of different potentials. The pressures used in a single RENS run are shown in sets. Tabulated iteration counts are shown for 10 GPa (NS) or the first set of pressures (RENS), but this increases with pressure.}
\label{ns_params_table}
\begin{tabular}{ ccccccc } 
\hline\hline
Potential & NS Algorithm & $P$ / GPa & $N$ & $K$ & $L$ & Iterations \\
\hline
Pb-EDDP & NS & 0.1, 5, 10, 15, 20, 30, 40, 60 & 30 & 480 & 600 & 325,000 \\ 
Pb-EAM & NS & 0.1, 1, 5, 10, 20, 30, 40, 50, 60 & 60 & 1200 & 1000 & 1,677,000 \\ 
Pb-MEAM & NS & 0.1, 1, 5, 10, 15, 20, 25, 30, 35, 40 & 60 & 1200 & 1000 & 1,668,000 \\ 
Pb-EDDP & RENS & \{1, 4, 7\},\{10, 13, 16, 19\} & 30 & 480 & 600 & 153,000 \\ 
Pb-EAM & RENS & \begin{tabular}{@{}c@{}}\{20, 25, 30, 35, 40, 45, 50\}, \{50, 52, 54, 56, 58, 60\}, \\ \{54, 56, 58, 60, 62, 64\}\end{tabular} & 60 & 400 & 1000 & 305,000 \\ 
Pb-MEAM & RENS & \begin{tabular}{@{}c@{}}\{5, 10, 15, 20, 25, 30, 35, 40\}, \\ \{30.0, 32.5, 35.0, 37.5, 40.0\}\end{tabular} & 60 & 400 & 1000 & 307,000 \\ 
\hline\hline
\end{tabular}
\end{center}
\end{table*}

Reference melting lines for FCC, HCP and BCC phases using the Pb-EAM model were determined using the {\tt calphy} Python library~\cite{Calphy} to perform thermodynamic integration calculations. Calculations were performed using 500 atoms and 15000 switching steps, with the experimental atmospheric melting temperature used as an initial starting point.

To support identification of crystalline phases, constant-pressure geometry optimisation was performed on saved NS configurations, from configurations representative of temperatures $10$~K below the melting points to the final configuration of the sampling. These optimisations were conducted in {\tt LAMMPS}, using a maximum force threshold of $2\times10^{-11} \mathrm{eV~\AA}^{-1}$. The potential energy landscape of Pb-EDDP appeared to be less smooth, so using the {\tt repose} energy minimiser~\cite{EDDPPackage} alongside {\tt LAMMPS} was necessary to avoid trapping in higher enthalpy local minima. 

Optimal crystal structures were also minimised across the whole pressure range studied, allowing free energy curves to be produced for each phase. This provided an estimate for where we expected solid-solid transitions to occur in the NS simulations. Cell symmetry was fixed so as to prevent higher energy configurations (such as low pressure BCC) optimising to other geometries.    

Steinhardt bond order parameters were used as a measure of local order~\cite{steinhardt_bond-orientational_1983}. The $Q_4$, $Q_6$, $W_4$ and $W_6$ parameters were calculated on a per-atom basis and averaged over each NS configuration, using the {\tt QUIP} package~\cite{Csanyi2007-py}, applying a cut-off radius of $4.0$~\AA~to include only the first neighbour shell. Values for the optimal FCC, HCP and BCC crystal structures that were used as reference are shown in Table \ref{steinhardt_bond_order_table}. The double hexagonal close-packed (dHCP) structure has values exactly halfway between FCC and HCP, representing its effective $1$:$1$ FCC:HCP stacking variant nature.

\begin{table}
\begin{center}
\caption{Steinhardt bond order parameter values for key crystal structures~\cite{steinhardt_bond-orientational_1983}.}
\label{steinhardt_bond_order_table}
\begin{tabular}{ cccccc } 
\hline\hline
\begin{tabular}{@{}c@{}}Crystal \\ Structure\end{tabular}& $Q_6$ & $W_6$ & $Q_4$ & $W_4$ & \begin{tabular}{@{}c@{}}Coordination \\ Number\end{tabular} \\
\hline
FCC & 0.5745 & -0.01316 & 0.1909 & -0.1593 & 12 \\ 
dHCP & 0.5293 & -0.01283 & 0.1428 & -0.01253 & 12 \\ 
HCP & 0.4842 & -0.01250 & 0.09467 & 0.1341 & 12 \\
BCC & 0.6285 & 0.01316 & 0.5092 & -0.1593 & 8\\ 
BCC & 0.5107 & 0.01316 & 0.03637 & 0.1593 & 14\\ 
\hline\hline
\end{tabular}
\end{center}
\end{table}

% \begin{table}
% \begin{center}
% \caption{Parameters used for the replica-exchange nested sampling calculations of different potentials. Each row represents a separate RENS run and its associated pressures.}
% \label{rens_params_table}
% \begin{tabular}{ cccccc } 
% \hline\hline
% Potential & $P$ / GPa & $N$ & $K$ & $L$ & Iterations \\
% \hline
% EDDP & 1, 4, 7 & 30 & 480 & 600 & 153,000 \\ 
% EDDP & 10, 13, 16, 19 & 30 & 480 & 600 & 170,000 \\
% EAM & 20, 25, 30, 35, 40, 45, 50 & 60 & 400 & 1000 & 305,000 \\ 
% EAM & 50, 52, 54, 56, 58, 60 & 60 & 400 & 1000 & 305,000 \\ 
% EAM & 54, 56, 58, 60, 62, 64 & 60 & 400 & 1000 & 307,000 \\ 
% MEAM & 5, 10, 15, 20, 25, 30, 35, 40 & 60 & 400 & 1000 & 307,000 \\ 
% \hline\hline
% \end{tabular}
% \end{center}
% \end{table}

%%%%%%%%%%%%%%%%%%%%%%%%%%%%%%%%%%%%%%%%%%%%%
%%%%%%%%%%%%%%%%%%%%%%%%%%%%%%%%%%%%%%%%%%%%%
%%%%%%%%%%%%%%%
\section{Results} \label{sct:results}

In this section, we analyse the thermodynamic properties derived from the simulations, examine the identification of solid configurations -- especially at high temperatures -- and consider the effects of the finite system size. We then present the resulting pressure–temperature phase diagrams for all three models, with particular attention to differences between NS and RENS.

%%%%%%%%%%%%%%%%%%%%%%%%%%%%%%%%%%%%%%%%%%%%%

\subsection{Thermodynamic properties}
% Density, enthalpy and heat capacity comparisons for all three models

\begin{figure*}[hbt]
    \centering
    \includegraphics[width=\linewidth]{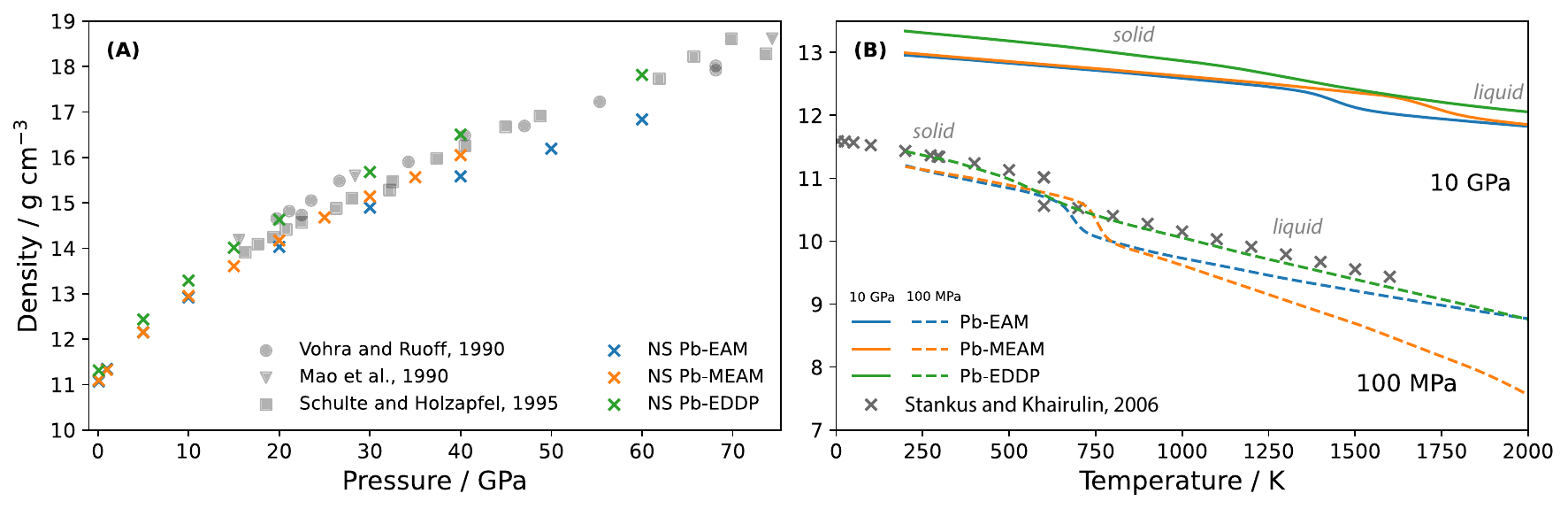}
    \caption{(A) Density of solid lead at $300$~K, as a function of pressure. Grey symbols show experimental results from Vohra and Ruoff~\cite{vohra_static_1990}, Mao~\cite{mao_high-pressure_1990}, and Schulte and Holzapfel~\cite{schulte_equation--state_1995}. Coloured crosses are densities calculated from the NS calculations for the three different studied potentials. (B) Density of lead at constant pressure. Grey crosses show experimental results from Stankus and Khairulin~\cite{stankus_density_2006}, obtained at atmospheric pressure. Lines show the density from NS calculations at $100$~MPa (dashed lines) and $10$~GPa (solid lines) for the three tested potential models.}
    \label{fig:high_p_density_fig}
\end{figure*}

We first calculated the expected values of density over a wide range of temperatures and pressures from the NS results for the three potential models considered. 
Fig.~\ref{fig:high_p_density_fig} summarises this comparison, as well as available experimental data. Panel A presents the predicted densities of solid lead at $T=300~\mathrm{K}$ as a function of pressure, showing good agreement overall with experiment, with the variance between the models comparable to the uncertainty of experimentally-measured values. Among the potentials benchmarked here, Pb-EDDP predicts the highest density across the entire pressure range and is in very good agreement with experimental results from Ref.~\cite{mao_high-pressure_1990}. Pb-EAM and Pb-MEAM density predictions are very close at low pressure, with Pb-EAM showing slightly lower densities as the pressure increases, with the difference between Pb-EDDP and Pb-EAM becoming $1.0\text{~g~cm}^{-3}$ at $40$~GPa.  

Fig.~\ref{fig:high_p_density_fig}B compares the predicted density across the solid and liquid phases up to $T=3000~\mathrm{K}$ at two different pressure values, $100$~MPa and $10$~GPa. The experimental results were obtained at atmospheric pressure.  
The density of the low-pressure solid is reproduced well overall by all three potentials, with the Pb-EDDP model being closest to experimental values. While Pb-EDDP follows the liquid density most closely, the Pb-EAM model also reproduces the gradient remarkably, as well as the density difference between the solid and liquid phases. 
At $10$~GPa, the density is affected significantly less by temperature, with all three models predicting similar trends.
% Based on this, the NS density aligns well across all reported temperatures for the EDDP, whereas Pb-EAM and Pb-MEAM notably predict lower liquid density and reasonable solid density.

%\textcolor{blue}{TODO: Enthalpy discussion....
%Enthalpy change of formation}

\begin{figure*}[htb]
    \centering
    \includegraphics[width=\linewidth]{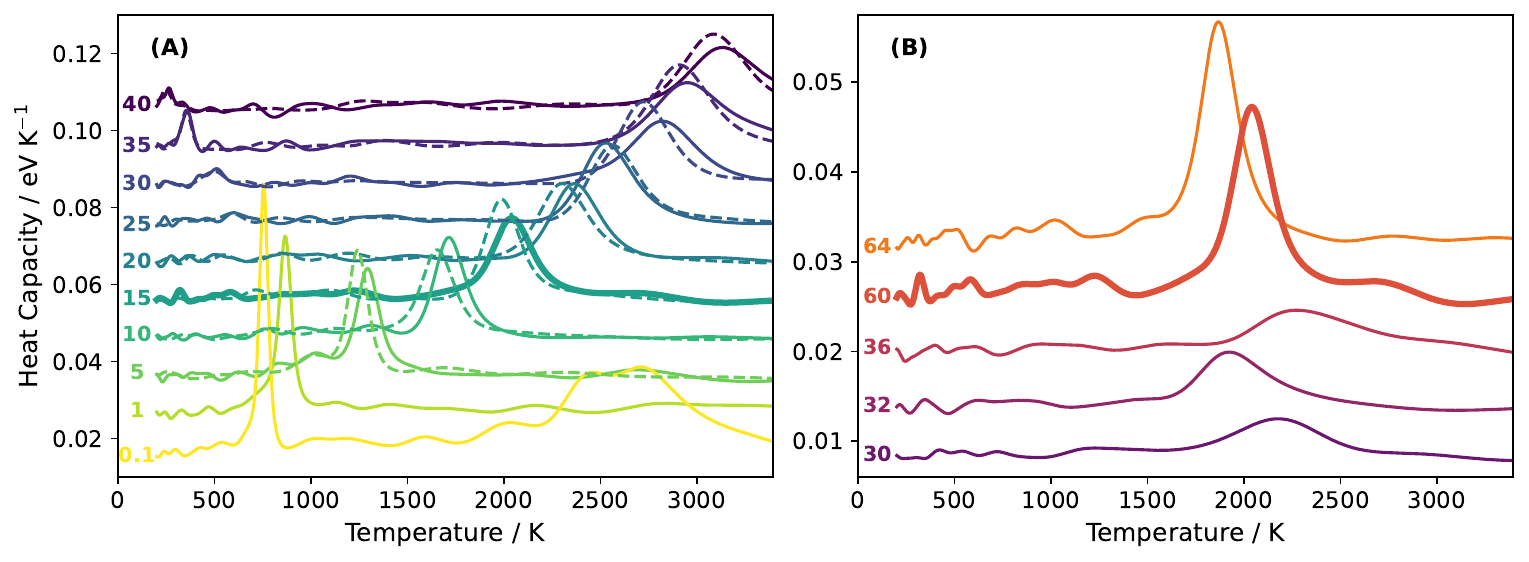}
    \caption{Constant-pressure heat capacity of the simulated systems as a function of temperature, using the Pb-MEAM model. Solid and dashed lines represent calculations performed using standard NS and RENS, respectively. (A) Different pressures, with values displayed in GPa by each curve, with a $60$-particle system size. Heat capacity values are shifted up by $0.01\text{~eV~K}^{-1}$ for each increasing pressure, to assist readability. (B) Different system sizes (with atom counts shown by each curve) at $15$~GPa. Heat capacity is shifted as a function of system size (increased by $0.005\text{~eV~K}^{-1}$ for each successive system size). The $15$ GPa, $60$-atom NS curves in each panel (representing the same run) are shown with a bolder line.}
    \label{fig:meam_heat_capacity}
\end{figure*}

We also evaluated the constant-pressure heat capacity from the NS calculations. While in the thermodynamic limit the heat capacity becomes infinite at first-order phase transitions, a broader Gaussian-like peak is observed instead in practice, due to the finite size of the system in the calculations.
As an example, Fig.~\ref{fig:meam_heat_capacity} displays the heat capacity as a function of temperature across the range of studied pressures for the Pb-MEAM model, evaluated both from NS and RENS calculations. The solid-liquid transition can be clearly identified and the melting temperatures determined by the location of the maximum in each curve, which increases with pressure, as expected~\cite{errandonea2010melting}. In addition, a further peak is visible for $30-40$~GPa simulations (with much smaller but still notable local maxima in the heat capacity), occurring at $510$~K, $360$~K and $265$~K respectively, suggesting a solid-solid transition in the system. 
At the lowest-pressure sampling, $p=100~\mathrm{MPa}$, a broad and shallow peak can be observed at high-temperature. This heat capacity maximum in the supercritical fluid region signals a point in the respective Widom-line.
Performing a series of NS calculations in the low-pressure regime and calculating the volume distribution at temperatures around the heat capacity peak would allow us to locate the critical point~\cite{pt_phase_dias_ns}. However, this single result at $100~\mathrm{MPa}$ already suggests that Pb-MEAM significantly underestimates the experimentally predicted critical pressure and temperature ($250\pm30~\mathrm{MPa}$ and $5400\pm400~\mathrm{K} $~\cite{ternovoi1996experimental}), and thus we did not investigate further. 

To demonstrate the effect of system size on the width of the heat capacity peaks and the melting temperature, we compare the results of NS runs with the Pb-MEAM potential at $p=15~\mathrm{GPa}$, using different numbers of atoms.
As seen in panel B of Fig.~\ref{fig:meam_heat_capacity}, 
the heat capacity peak generally becomes more pronounced and narrower with system size, with smaller systems typically overestimating the melting temperature. However, the changes allowed to the shape of the cell during the exploration, and the way various crystalline structures can be accommodated, will influence this trend~\cite{NS_mat_review}. One should also note that distinguishing peaks corresponding to phase transitions from those of fluctuations caused by the finite resolution of the sampling becomes harder for smaller systems, particularly in the case of solid-solid transitions. While the RENS approach can reduce fluctuations in thermodynamic observables successfully, it often becomes necessary to use structural descriptors to successfully locate transitions, especially when computational cost -- such as that of the Pb-EDDP model -- limits the system's size. 

%%%%%%%%%%%%%%%%%%%%%%%%%%%%%%%%%%%%%%%%%%%%%

\subsection{Identifying solid structures}
% Discuss possible present phases and how finite temperature effects their identification. 

%\sh{Needs an introduction paragraph}
Reliable structural identification is essential for locating phase transitions and determining thermodynamically relevant solid phases, for example by allowing us to calculate the partition function and, hence, the free energy of the basin of attraction of a given solid phase separately. 

\subsubsection{Stacking variants at finite temperature}

\begin{figure}[hbt]
% Stacking variant illustration
    \centering
    \includegraphics[width=\linewidth]{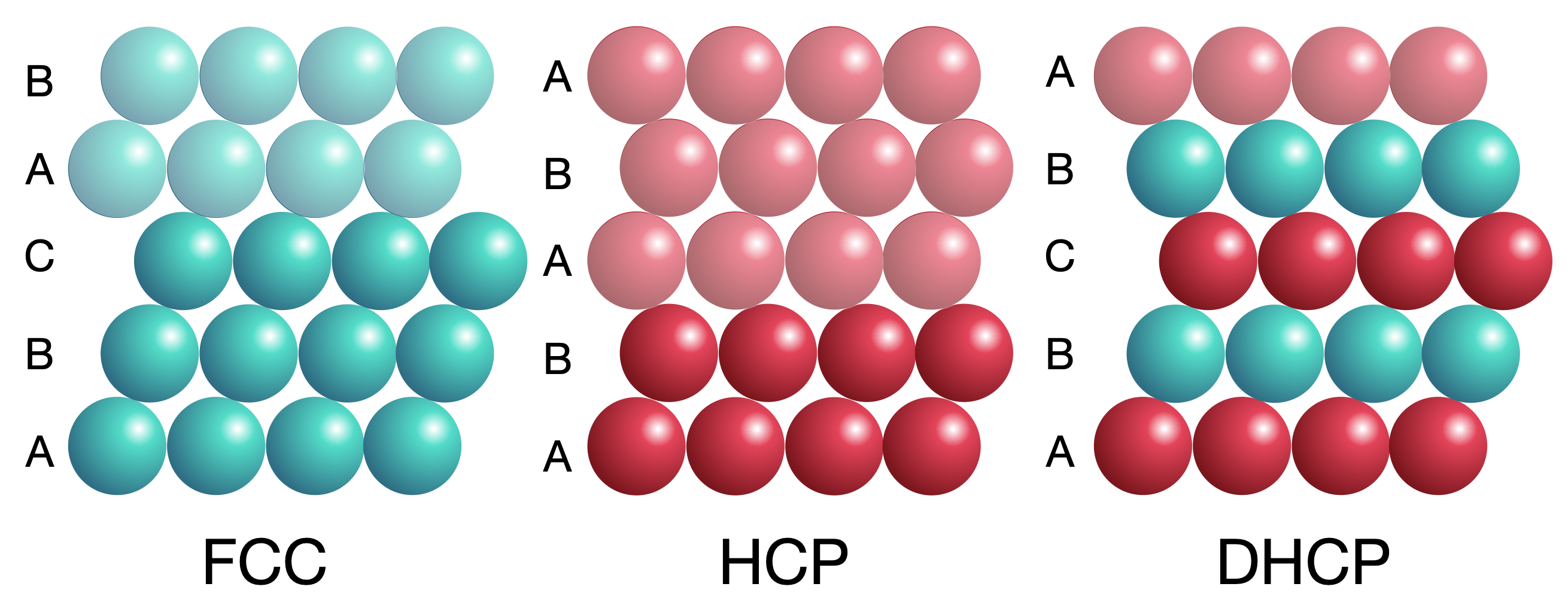}
    \caption{Schematic illustration of polytype face-centred cubic (FCC), hexagonal close-packed (HCP), and double hexagonal close-packed (dHCP) structures. ABC notation on the left marks the relative stacking position, with layers sandwiched by two different stackings in blue, and layers sandwiched by two sheets of atoms in the same relative position coloured red.}
    \label{fig:stacking_demo}
\end{figure}

Low-temperature solid configurations can usually be readily identified and categorised using order parameters based on local structure and symmetry. However, our top-to-bottom sampling from the liquid phase towards crystalline configurations necessarily produces high-temperature solid samples, in which thermal fluctuations blur crystalline motifs, complicating straightforward categorisation. 
The situation is further complicated for close-packed structures, where a wide range of stacking variants can occur. To aid the following discussion, we illustrate the FCC, HCP and dHCP structures in Fig.~\ref{fig:stacking_demo}, highlighting that they are polytypes built from the same close-packed hexagonal layers but stacked in different sequences. In principle, an infinite number of stacking variants can be constructed, corresponding to arbitrary combinations of FCC-like and HCP-like local stacking. Aside from the aforementioned close-packed variants, the BCC structure is also relevant for the studied lead system. We have not found other crystalline structures in any of the simulations. 

Steinhardt bond-order parameters~\cite{steinhardt_bond-orientational_1983} can in principle distinguish between these structures (see Table~\ref{steinhardt_bond_order_table} for values in perfect crystals), and we therefore use them as the basis of structural identification. For each configuration generated during NS, we calculate the average Steinhardt parameters and compare them with the values for the reference crystalline phases. In finite-temperature solids, however, thermal fluctuations cause deviations from the ideal values -- an effect that becomes increasingly pronounced at higher temperatures, particularly near or above the barriers separating competing structures, making it difficult or even impossible to reliably classify them.

\begin{figure*}
    \centering
    \includegraphics[width=\linewidth]{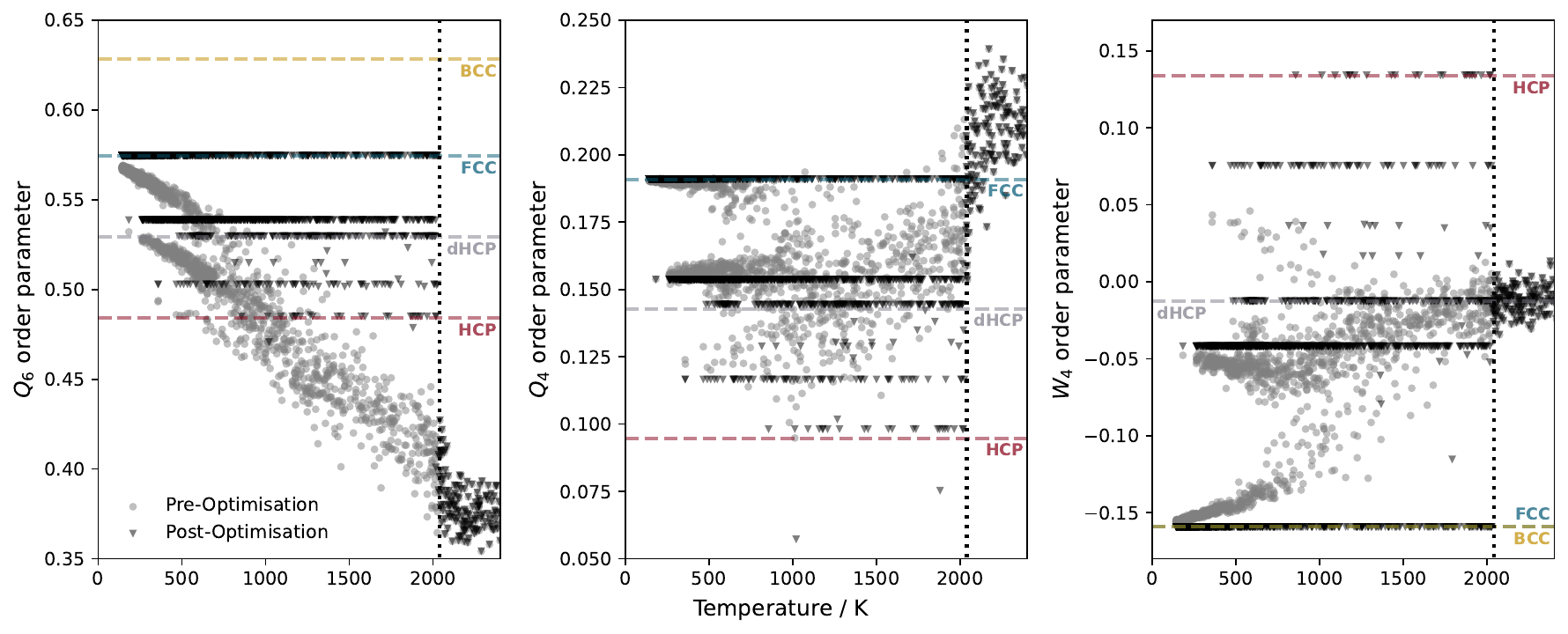}
    \caption{Configuration-averaged Steinhardt bond order parameters of an exemplar NS run ($15$ GPa, Pb-MEAM, $60$ atoms). Values for raw NS configurations are shown by grey circles, while the order parameter values after the geometry optimisation are shown by black triangles. Optimal crystal structure values for FCC, HCP, dHCP and BCC are shown by horizontal dashed lines. The melting temperature at this pressure is marked with a vertical dotted line.}
    \label{fig:ops_comparison}
\end{figure*}

To overcome this challenge, we performed constant-pressure geometry optimisations of all solid configurations generated during the sampling, allowing them to quench into the nearest local enthalpy basin. 
Fig.~\ref{fig:ops_comparison} illustrates this through the example of the evolution of order parameters $Q_6$, $Q_4$ and $W_4$ as a function of temperature for configurations sampled by NS at $15$~GPa using Pb-MEAM and 60 atoms, both before and after the optimisation step. 
These figures also provide a visual impression of free energy differences, as the relative ratio of configurations associated with the basins of attraction of different structures represents their relative phase space volume.
At higher temperatures, the order parameters associated with the raw NS output configurations are obfuscated. However, these values clearly converge to specific order parameter values at low temperature, making it possible to distinguish at least three separate basins in this case.

In contrast, the order parameter values for the optimised structures are distinct, as almost all configurations have converged to a crystalline structure: FCC, dHCP, HCP and two other structures with order parameters that sit in between the above expected values. These represent polytype structures, firstly where the ratio of FCC and HCP layers is $3$:$2$, observed in two different stacking variants [ABCBACABACBCACB]$_n$ (often referred to as the 15R structure) and [ABACB]$_n$, and secondly where the ratio of FCC and HCP layers is $1$:$4$, [ABCABCBACB]$_n$.

While identifying FCC and a stacking variant as the most dominant structures in the system is also possible from the raw NS configurations, distinguishing higher temperature configurations, particularly stacking variants, becomes possible only with the optimised structures. 
We thus used this workflow to identify and categorise solid phase configurations for free energy calculations and determining the phase diagram.  
Although this procedure greatly reduces the variance of chosen order parameter values and thus makes structural assignment easier, it relies on the assumption that quenching is unique and preserves the relative phase space volume ratio of the basins of attraction.
Furthermore, it has to be noted that finite temperature dynamically-stable structures cannot be captured via quenching, although we have not found any evidence of such structures within the studied lead systems.

We also have to note that, in some cases, high temperature structures did not optimise to a recognisable crystalline structure. Of particular note is that many structures optimised with the Pb-EDDP potential in {\tt LAMMPS} had a small number of atoms with non-physical short interatomic distances in the range of $1.3$~\AA~to $1.7$~\AA. These are likely associated with a `hole' in the potential energy surface -- an energy basin, where a non-physical structure is calculated to have a particularly low energy compared to realistic structures, though in this case these structures represent only local minima. Furthermore, given the absence of these structures in the NS simulations, these holes must have a very small phase space volume and are thus thermodynamically irrelevant. However, when assigning individual configurations to phases, this can present challenges. We also found that repeating the optimisation process with the {\tt repose} energy minimiser resulted in fewer non-physical structures being produced. 

\subsubsection{Finite size effect on stacking variants}

While a range of stacking variants is observed, particularly in the high-temperature solid region, it is important to note that both the finite size of the system and the constraints on lattice distortions influence which stacking sequences can be realised. Certain polytypes are only compatible with specific periodicities; for example, the 15R structure described above requires at least a five-layer repeat unit depending on the permitted tilt of the cell and can therefore only be accommodated when the system size is commensurate with this periodicity, for example with a $60$-atom cell, but not when using $64$ atoms.
In addition, small simulation cells may favour a broader distribution of stacking sequences. This can arise from finite-size effects, including the relative contribution of configurational entropy associated with stacking disorder~\cite{chen2021thermodynamically, jiang2025role, mau1999stacking}. As the number of particles increases, these effects are expected to diminish, and the relative stability of different polytypes becomes increasingly governed by their energetic differences, although this balance may remain sensitive when these differences are small. Nevertheless, defects such as stacking faults can remain relevant at high temperatures, and the ability of nested sampling to capture such configurations is valuable, as it provides access to experimentally relevant structures.

%Given the presence of stacking variants between FCC and HCP, we investigated the impact of system size on their occurrence. For instance, the aforementioned 15R structure periodically repeats across 5 atoms. With a 64-atom unit cell, which cannot be factorised to have a 5-atom length in any dimension, this structure cannot be produced. As such, we can consider this structure to be an artefact of the system size. Other system sizes may also produce artificial configurations due to their size constraints, so a direct comparison was performed for 30, 32, 36, 60 and 64-atom cells, using the Pb-MEAM potential at 15 GPa. 

\begin{figure}
    \centering
    \includegraphics[width=\linewidth]{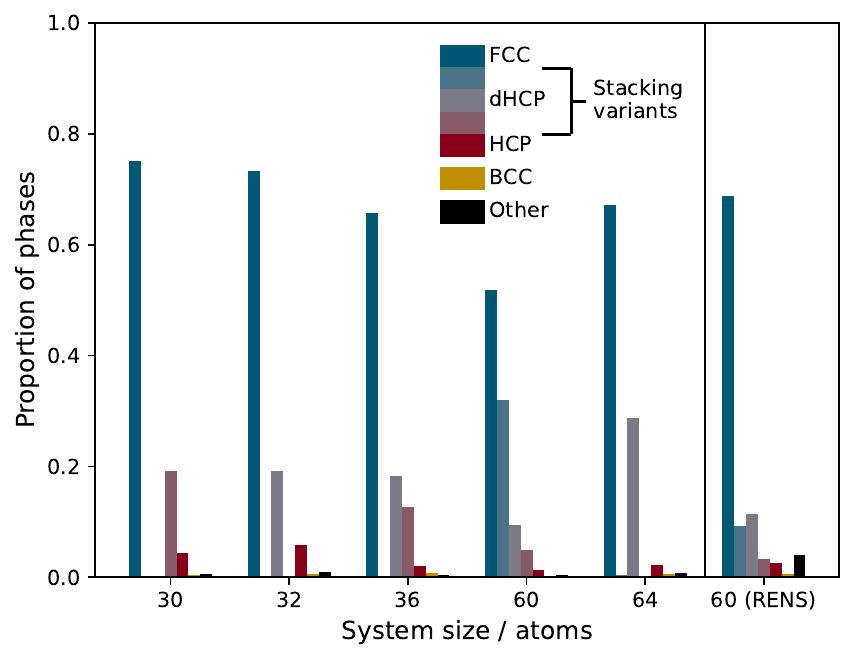}
    \caption{Relative proportions of FCC, HCP, BCC and stacking variants observed for different system sizes in $15$ GPa MEAM NS runs and for a $60$-atom RENS calculation.}
    \label{fig:system_sizes}
\end{figure}

In this section, we therefore examine the distribution of stacking sequences as a function of system size and compare findings for $30$, $32$, $36$, $60$ and $64$-atom cells, using the Pb-MEAM potential at $15$~GPa.
Figure~\ref{fig:system_sizes} depicts the proportion of structures observed after enthalpy minimisation across the solid region of each of these NS runs, with configurations that cannot be categorised into any of the crystalline phases shown in black. In this case, all system sizes were dominated by the thermodynamically stable FCC structure, with only a negligible number of HCP configurations. 
The $32$ and $64$-atom cells only featured the dHCP structure apart from these, while the $30$-atom formed the 9R structure instead, with $36$-atom simulations sampling both of these stacking variants in low numbers. Most importantly, although which specific stacking variant is realised in these simulations is different -- depending on what can be accommodated in the cell -- their relative proportion in the sampling is roughly constant, showing the size of the system has only a minor effect on the exploration of the solid phase. 
Results obtained with the $60$-atom NS are somewhat different, having the widest variety and highest proportion of stacking variants. Since a $60$-atom triclinic cell can accommodate the close-packed solid with different number of layers, observing a wider range of polytype structures is not surprising. A more detailed analysis of the data also shows that this increase in the proportion of polytypes is highest within $100$~K of the melting line, and as the temperature decreases, the proportion of stacking variants becomes even more similar across different system sizes. 
For comparison, results from the $15$~GPa replica of a $60$-atom RENS simulation is also shown. This demonstrates our overall finding across the models that RENS samples fewer polytypes, but in more even distribution than NS. However, the number of configurations that remain amorphous after the geometry optimisation is higher in the case of RENS, most likely as a result of solid and liquid-like configuration exchanges between replicas near the melting point. 
Overall, these results suggest that as long as the thermodynamically relevant crystalline phases can be accommodated in the simulation box, the precise number of atoms has a limited effect, even in these very small systems.
To maximise flexibility and variability, the $60$-atom system was selected for the sampling of the Pb-EAM and Pb-MEAM phase diagrams. However, due to the computational expense of evaluating the Pb-EDDP model, the compromise of using $30$ atoms had to be made. %Based on this system size comparison, the 30-atom cell should be less problematic than 60 atoms regardless, though at the cost of potentially not capturing any structures with larger unit cells. In any case, such structure were neither expected nor observed in this work. 

%%%%% Enthalpy curve plots %%%%%
\subsubsection{Ground state structure}

\begin{figure}[h]
    \centering
    \includegraphics[width=\linewidth]{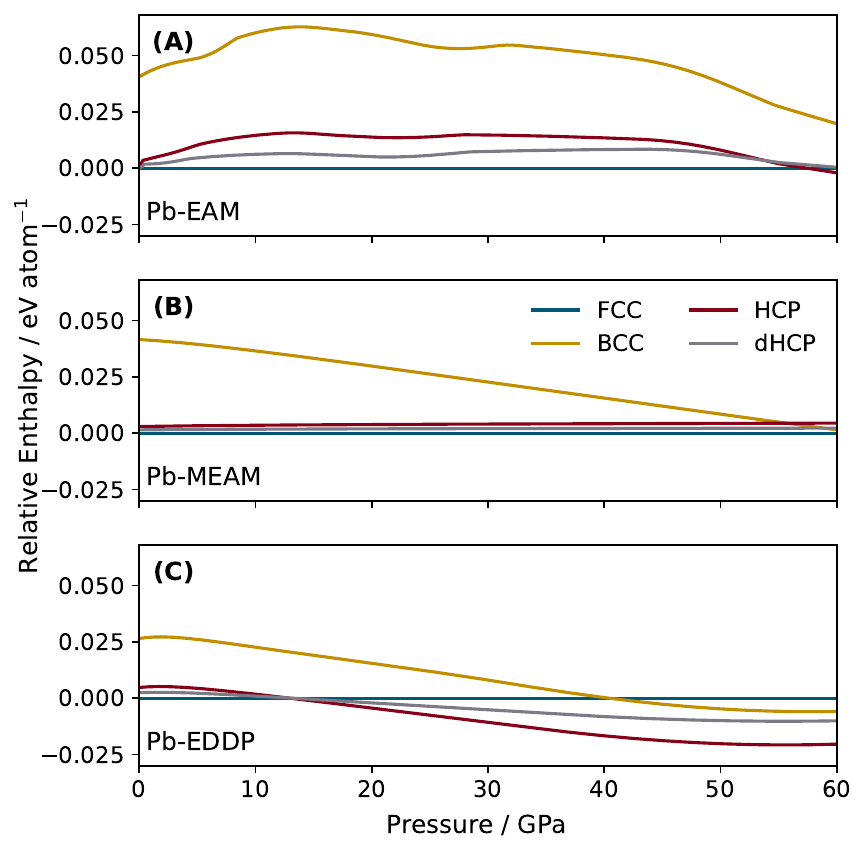}
    \caption{Enthalpy difference of selected crystal structures of lead, relative to FCC, calculated performing geometry optimisation using (A) Pb-EAM, (B) Pb-MEAM and (C) Pb-EDDP models across the studied pressure range.}
    \label{fig:enthalpy_curves}
\end{figure}

To determine the ground state structures as a function of pressure, we performed constant-pressure enthalpy minimisation of FCC, HCP, dHCP and BCC structures. Fig.~\ref{fig:enthalpy_curves} shows the resulting minimum enthalpy values relative to the FCC structure. All three studied models predict that the FCC structure is the global enthalpy minimum at low pressure, in agreement with experimental findings. However, they vary significantly in the predictions of their higher pressure behaviour.  

The Pb-EAM results predict that the FCC phase remains the ground state up to $57.5$~GPa, when a transition to the HCP phase occurs. This contradicts the original publication of the model describing the FCC to HCP transition at $13$~GPa. Since we have used the model through the available {\tt LAMMPS} potential file~\cite{PotentialDatabase}, it is possible that differences in how the potential file was created or tabulated caused this discrepancy.

The Pb-MEAM model predicts an FCC ground state for the studied pressure range, with enthalpy trends suggesting an FCC-BCC transition by $60.5$~GPa, while the HCP phase does not become the global minimum in the studied pressure range.
The Pb-EDDP predictions are overall closest to expectations. FCC is the lowest enthalpy phase up to $13.0$~GPa, at which point the HCP phase becomes most stable for the remainder of the studied region.

%%%%% PHASE DIAGRAMS %%%%%

\subsection{Phase diagrams}

The pressure-temperature phase diagrams were constructed using the information from individual NS calculations, or a series of RENS calculations performed at different pressures. 
Melting points were determined by the location of heat capacity peaks and error bars correspond to the full width at half maximum (FWHM), to indicate the widening of the heat capacity peak caused by the finite size of the simulated systems. On the phase diagrams shown here, we also indicate which structures were observed during the sampling. These coloured bars are constructed from individual horizontal segments, each segment representing a single configuration and coloured according to the basin of attraction determined by geometry optimisation: blue for FCC, dark red for HCP, and intermediate shades representing stacking variants of various ratios of FCC-like and HCP-like layers. BCC structures are represented with yellow, and those (primarily high temperature) configurations that failed to minimise to a crystalline state are shown in white.
Random samples below the melting point were selected, representing $0.4$~\% of all NS configurations. This equates to between $3000$ and $6000$ included configurations per pressure for each model -- the number generally increasing as the melting temperature increases with pressure.
The temperature associated with each configuration represents the point where that configuration has the highest probability. This was determined after the sampling was finished, selecting the temperature where the expected value of the enthalpy (calculated from the partition function) is equal to the individual enthalpy of a configuration.

%%%%%%%%%%%%%%%%%%%%%%%%%%%%%%%%%%%%%%%%%%%%%
\subsubsection{Pb-EDDP phase diagram}

\begin{figure}
    \centering
    \includegraphics[width=\linewidth]{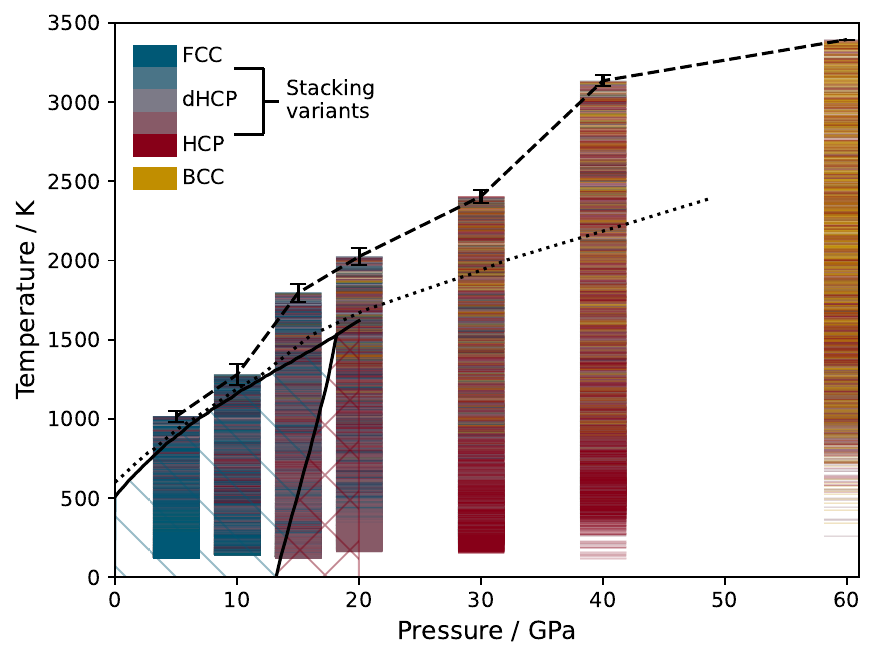}
    \includegraphics[width=\linewidth]{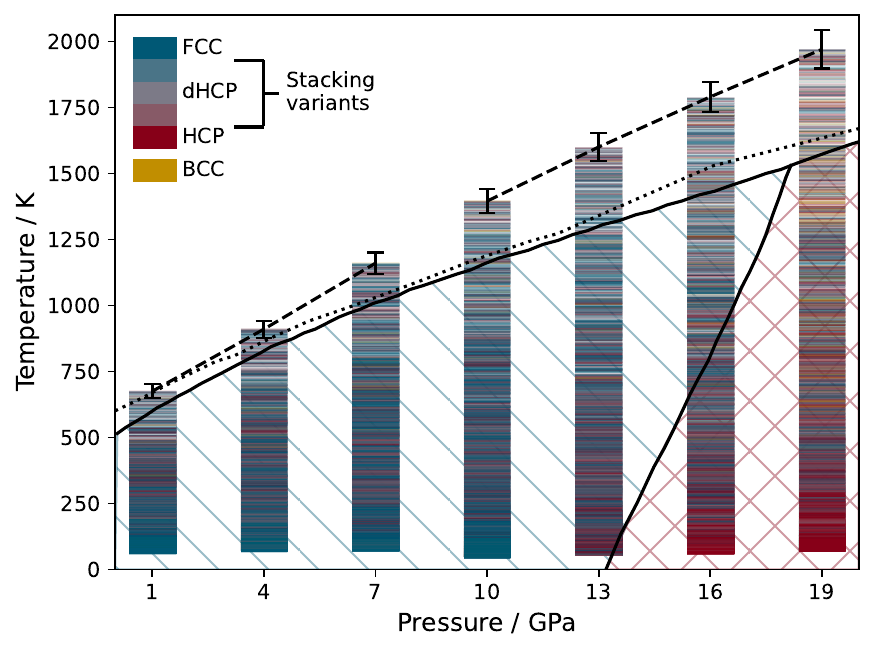}
    \caption{Pressure-temperature phase diagram of the Pb-EDDP model based on NS (top panel) and RENS (bottom panel) calculations. Black dashed lines show the melting line determined in the current work, black dotted lines show experimental melting temperatures determined by Errandonea up to $12$~GPa~\cite{errandonea2010melting} and Godwal {\it et al.} beyond this~\cite{godwal_ultrahigh-pressure_1990}. Solid black lines show phase transition information for Pb-EDDP from Ref.~\cite{salzbrenner2023}, with blue and red hashed regions representing FCC and HCP phases, respectively. Coloured bars represent structures sampled during NS and RENS. }
    \label{fig:eddp_phase_diagram}
\end{figure}

Figure~\ref{fig:eddp_phase_diagram} shows the phase diagram of the Pb-EDDP model calculated by NS and RENS. Previously reported Pb-EDDP melting and solid-solid transition lines determined using coexistence MD with spin-orbit coupling are shown in black~\cite{salzbrenner2023}, with hashed backgrounds representing the stable phases (FCC and HCP) in each region. Experimental melting temperature data are also shown for reference~\cite{errandonea2010melting,godwal_ultrahigh-pressure_1990}.
The NS melting line aligns reasonably well with these previous results at lower pressures, overestimating the melting temperature by less than $100$~K due to the finite system size. This deviation from prior results increases at higher pressures. 
As expected, walker exchanges between the consecutive pressures in RENS make the sampling smoother, with more even increments in the melting line. 

Stacking variants between FCC and HCP structures were commonly observed at higher temperatures, but at different proportions depending on the pressure. For example, stacking variants [ABCACB]$_n$ (also referred to as triple-HCP or 6H$_1$) and [ABACACBCB]$_n$ (also referred to as 9R), represented $2$:$1$ and $1$:$2$ ratios of FCC and HCP layers, respectively. One has to note though, that further stacking variants cannot be accommodated using $30$ atoms, unless significantly more distortion of the cell is allowed. 

At $5$~GPa, the sampling is dominated by FCC configurations in the NS calculations, with a small amount of HCP configurations appearing at $10$~GPa. A high proportion of stacking variants was observed around the transition region, $15$ and $20$~GPa -- the latter of which has almost no remaining FCC configurations. By $30$~GPa, the dominant phase is HCP, with trace amounts of BCC at higher temperatures, while BCC represents around $52$~\% of all solid configurations at $60$~GPa.

\begin{figure}[h]
    \centering
    \includegraphics[width=\linewidth]{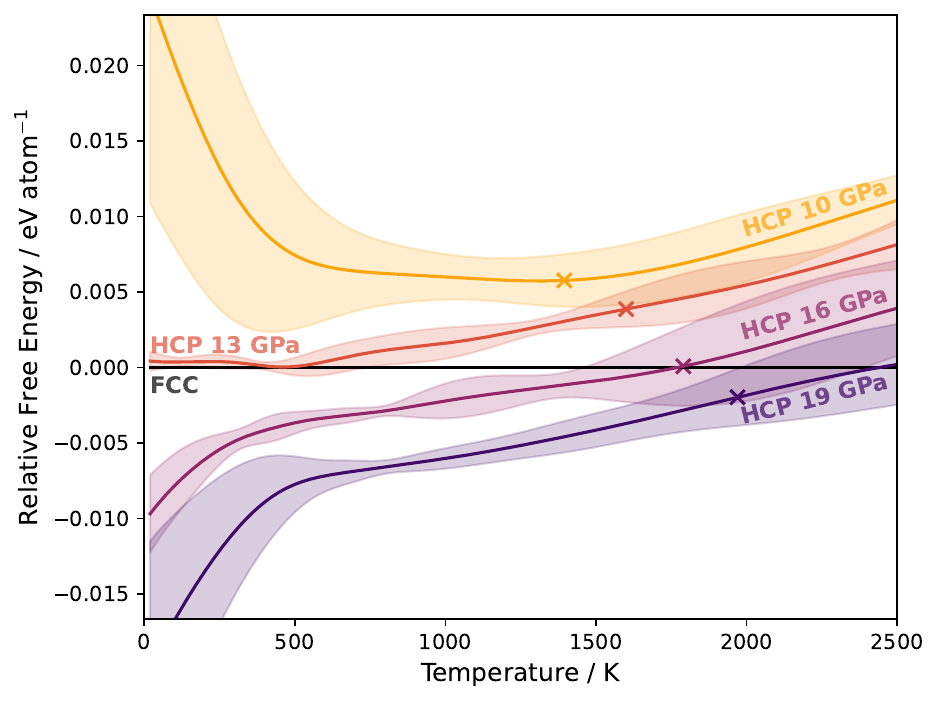}
    \caption{Gibbs free energy of the HCP phase relative to the FCC baseline at the same pressure around the solid-solid transition line, based on Pb-EDDP RENS calculations. Crosses represent the melting temperatures at each corresponding pressure. Shaded regions represent the error in the form of two standard deviations from the mean, based on eight calculations each representing $0.1$~\% of RENS configurations.}
    \label{fig:fcc_hcp_free_energies}
\end{figure}

In contrast to the NS calculations, RENS presents fewer stacking variants at higher temperatures below the melting line (but more configurations that were not minimised to any of the studied crystalline structures), probably due to swapping configurations across the liquid-solid phase boundary between consecutive pressures. Furthermore, the anticipated solid-solid transition between $13$ and $19$~GPa is sampled more clearly, presenting as smoother transition from FCC to HCP. 
As the assignment of optimised structures to crystalline phases also enables the calculation of their relative free energies as a function of temperature, we can characterise this transition quantitatively. Figure~\ref{fig:fcc_hcp_free_energies} displays the Gibbs free energy of the HCP phase compared to that of the FCC phase (with contributions from other stacking variants not included) at pressures around the solid-solid transition line from RENS simulation of the Pb-EDDP potential. 
As it is also clear from the visual representation of the sampled configurations on the phase diagram (Fig.~\ref{fig:eddp_phase_diagram}), the FCC phase is the more stable in the entire temperature range below the melting line at $10$~GPa. 
At $13$~GPa the FCC phase still dominates at higher temperatures, but the free energy difference between the two structures decreases while approaching the phase boundary and becomes imperceptible below $500$~K. At $16$~GPa the HCP phase becomes more stable below $1730$~K, although this value is very close to the melting temperature determined from the RENS calculations ($1790$~K) and likely to be strongly influenced by how the configurations near the melting point are assigned to basins. This result nevertheless shows a slightly steeper FCC-HCP phase boundary than the original quasi-harmonic approximation calculations\cite{salzbrenner2023}, suggesting an even closer alignment with the experimental liquid-FCC-HCP triple point\cite{errandonea2010melting}. 
%A caveat of this approach is that the free energy curves are calculated under the aforementioned assumption of relative phase space volume of basins being preserved during the crystal structure optimisation. However, given that, with very few exceptions, the optimised structures followed expected trends in order parameters relative to the raw configurations, it is appropriate to consider these curves largely accurate. Furthermore, these curves do not factor in stacking variants -- given that these largely arise due to the limited cell size, they would likely contribute to the HCP and FCC curves at sufficiently larger system sizes.

%%%%%%%%%%%%%%%%%%%%%%%%%%%%%%%%%%%%%%%%%%%%%
\subsubsection{Pb-EAM phase diagram}

\begin{figure}
    \centering
    \includegraphics[width=\linewidth]{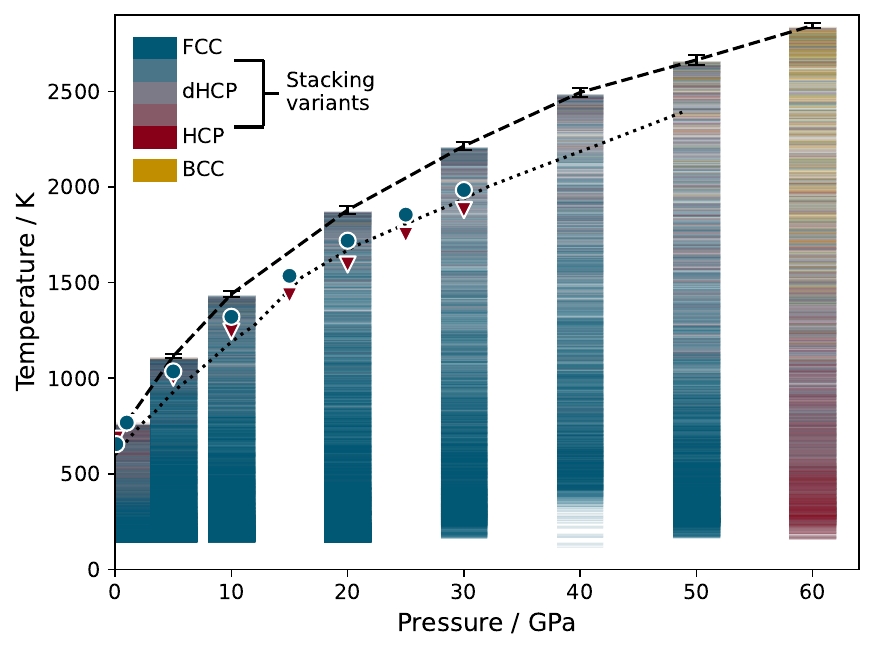}
    \includegraphics[width=\linewidth]{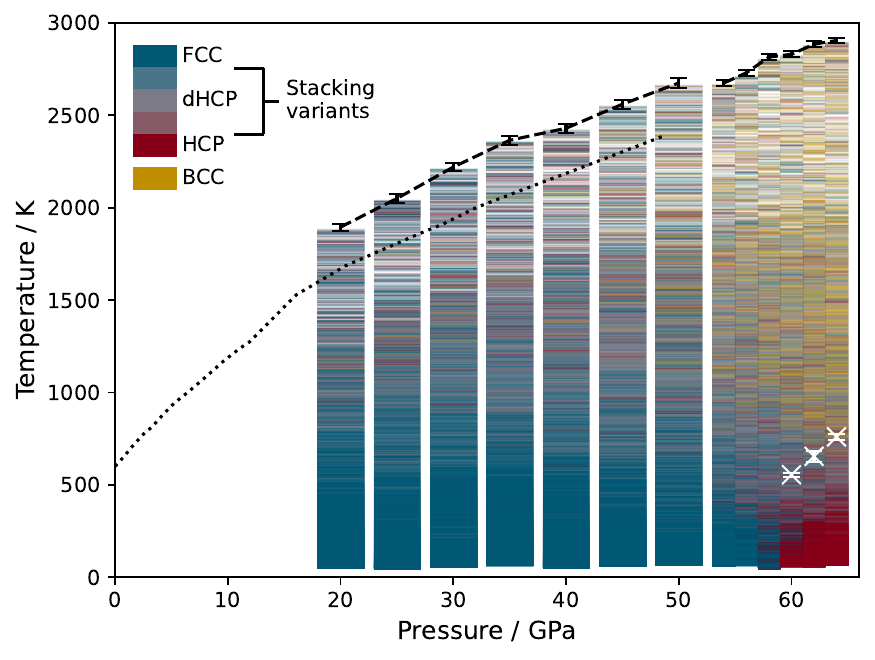}
    \caption{Pressure-temperature phase diagram of the Pb-EAM model based on NS (top panel) and RENS (bottom panel) calculations. Black dashed lines show the melting line determined in the current work, and black dotted lines show experimental melting temperatures determined by Errandonea up to $12$~GPa~\cite{errandonea2010melting} and Godwal {\it et al.} beyond this~\cite{godwal_ultrahigh-pressure_1990}. Blue circles and red triangles represent melting point values calculated using thermodynamic integration for the FCC and HCP phases, respectively. White crosses represent the solid-solid transition temperatures determined from the corresponding heat capacity peaks, with full width at half maximum (FWHM) error bars.}
    \label{fig:eam_phase_diagram}
\end{figure}

The Pb-EAM potential presents a very different phase behaviour, as seen in Fig.~\ref{fig:eam_phase_diagram}. 
The calculated melting line follows the expected trend closely. In the Pb-EAM case, we also determined the melting temperature using thermodynamic integration, which provides a useful reference for assessing finite-size effects in NS. As observed in previous systems\cite{NS_mat_review}, the $60$-atom NS simulations overestimate the melting temperature by approximately $8-10$~\%.
These reference calculations, performed using both FCC and HCP phases, also highlight the sensitivity of the predicted melting temperature to the choice of crystalline reference, with non-negligible differences between the two. In contrast, NS samples solid phases in an unbiased manner, avoiding the need to assume \textit{a priori} knowledge of the stable crystal structure.

Based on the NS calculations, the solid phase is strongly dominated by FCC up to $\approx50$~GPa, with only a small amount of stacking variants (mostly FCC-dominated) occurring, and primarily at higher temperatures.
At $60$~GPa, the ground state is found to be HCP, as expected; however, a wide variety of stacking variants and even some BCC configurations are observed. This is expected for Pb-EAM, as the enthalpy difference between polytype structures becomes smaller at higher pressures (see Fig.~\ref{fig:enthalpy_curves}), making entropic contributions more significant.
We further clarified the FCC-HCP transition using RENS calculations, with particular emphasis on the region between $54$ and $64$~GPa (Fig.~\ref{fig:eam_phase_diagram}, bottom panel). A fine pressure grid was chosen, to maintain high overlap between replicas and thus enable efficient walker exchange: swap acceptance rates were around $60$~\%, $45$~\% and $40$~\% at $3000$, $2000$ and $1000$~K, respectively, showing optimal sampling and convergence.
The results show that, while the proportion of stacking variants is higher at high temperature in the case of RENS, configurations below $58$~GPa were dominated by the FCC, and by the HCP structure above. Our free energy calculations show that at $58$ and $60$~GPa, the Gibbs energy difference between FCC and HCP is negligible in the entire temperature range. 
At higher pressures there is also a visible level of structures classed as BCC, although the free energy calculations confirm their contributions remain negligible compared to the close packed structures. 
Given the expected HCP to BCC transition at pressures above this (which would occur at approximately $90$~GPa, following on from the Pb-EAM enthalpy trends in Fig.~\ref{fig:enthalpy_curves}), it appears the BCC basin will continue to have larger phase space volume at higher pressures.

%FCC is also the dominant phase in the RENS phase diagram (Fig.~\ref{fig:eam_phase_diagram}). Even at 60 GPa, it remains present -- though there is a hint of HCP configurations at around 300 K to 400 K. Close to the melting line, HCP structures become more populous across all pressures. This is potentially due to the caveat of minimising structures first; configurations that have not become `trapped' in a single basin can conceivably be minimised into a number of structures.

%%%%%%%%%%%%%%%%%%%%%%%%%%%%%%%%%%%%%%%%%%%%%
\subsubsection{Pb-MEAM phase diagram}

\begin{figure}
    \centering
    \includegraphics[width=\linewidth]{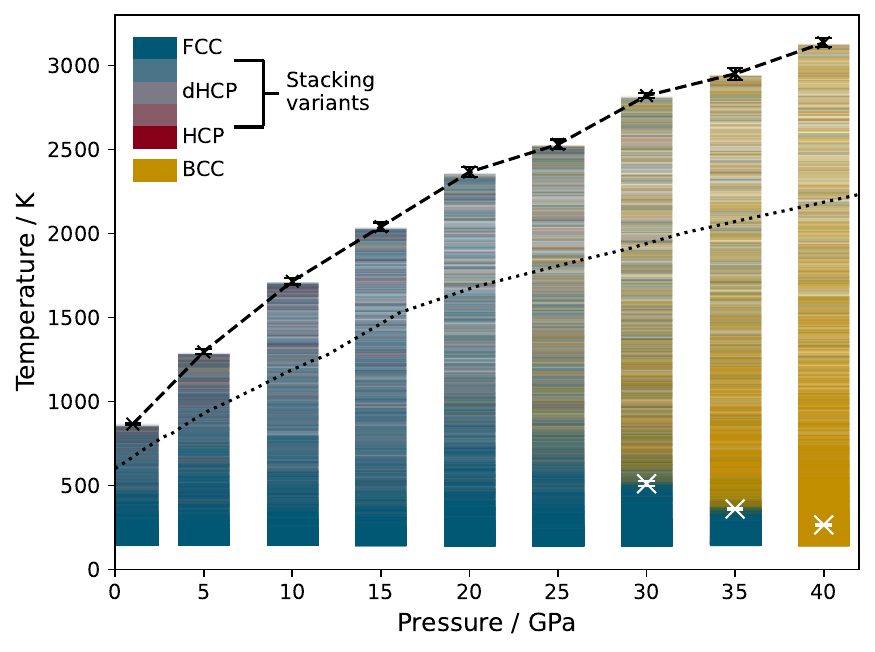}
    \includegraphics[width=\linewidth]{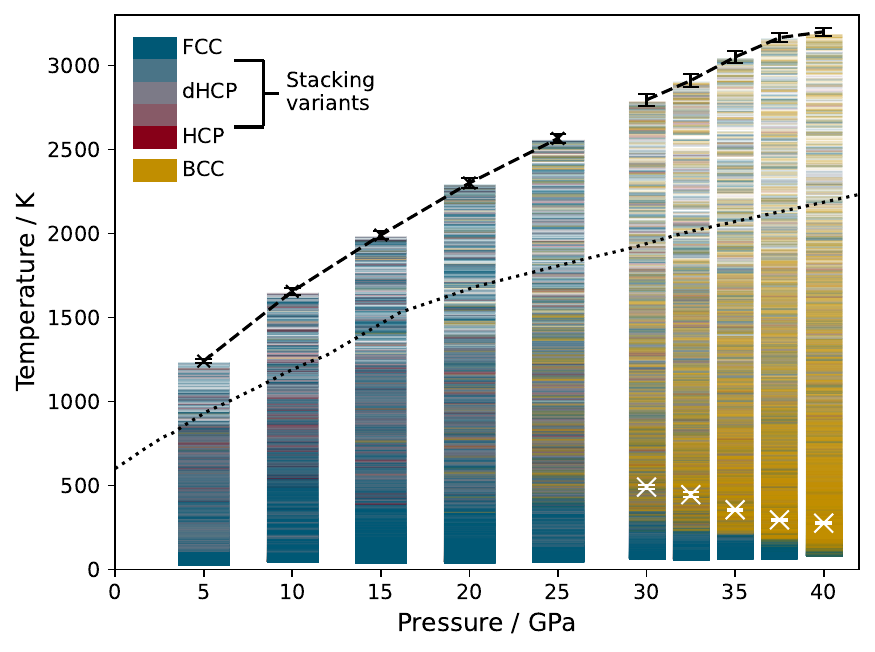}
        \caption{Pressure-temperature phase diagram of the Pb-MEAM model based on NS (top panel) and RENS (bottom panel) calculations. Black dashed lines show the melting line determined in the present work, and black dotted lines show experimental melting temperatures determined by Errandonea up to $12$~GPa~\cite{errandonea2010melting} and Godwal {\it et al.} beyond this~\cite{godwal_ultrahigh-pressure_1990}. White crosses represent the solid-solid transition temperatures determined from the heat capacity peaks in Fig. \ref{fig:meam_heat_capacity}, with full width at half maximum (FWHM) error bars.}
    \label{fig:meam_phase_diagram}
\end{figure}

The Pb-MEAM phase diagram, shown in Figure~\ref{fig:meam_phase_diagram}, paints a different picture yet again. The predicted melting temperatures are at least $200$~K higher across the entire studied range than those of the other two models, and the HCP structure is almost entirely absent.
While stacking variants are sampled near the melting line, these are all FCC-dominated, despite their relatively small enthalpy difference (see Fig.~\ref{fig:enthalpy_curves}). 
Instead, a very sharp transition to the BCC structure is observed at higher pressures, with a negative gradient due to the BCC phase being more dense than the FCC. The gradient of the transition line is also shallow, and we cannot make predictions about the location of the FCC-BCC-liquid triple point from the current results.
As the transition appears to be sharp (suggesting that free energy differences away from the coexistence point are significant), lower-resolution NS may struggle to retain the basin that is only relevant at low temperatures within the sampled configurations. As a consequence, the global minimum can be identified erroneously; for example, in the $40$~GPa NS run, the loss of all FCC configurations led to the system converging to the BCC phase (interestingly, a clear peak can still be identified on the heat capacity curve at $275$~K -- this is caused by a small but sharp increase in BCC density, rather than change in the atomic structure). Increasing the number of walkers can mitigate this issue, although a more computationally efficient solution is to employ RENS.

%The NS calculations found BCC becomes the global minimum by 40 GPa, given the complete lack of any FCC configurations (meaning there cannot be a transition to FCC at a temperature below the termination of calculations). Meanwhile, the RENS calculations still capture a BCC to FCC transition at this pressure, implying the global minimum does not become BCC until pressures exceeding the studied range. 

%\textit{Pascal S: I'd be really interested in some quantitative \& qualitative discussion of the relative strengths \& weaknesses of the EDDP \& (M)EAM for all the applications you present here. Clearly the EDDP is a lot more expensive, but how much exactly? I understand that as a result, we cannot sample configurations with as many atoms. How big a problem is this? Are finite size effects significant enough to outweigh the advantage of (in principle) near-ab initio accuracy? I suspect not from the phase diagrams, but it could be good to say this. How does using the EDDP compare to previous attempts at NS with MLIPs? What does the development of new generations of MLIPs mean for NS? I think sampling the partition function directly is kind of like a dream. Clearly the expense is a major problem, but how and how quickly are we approaching closer to a really comprehensive understanding of crystal phase diagrams using this method + MLIPs?}

%%%%%%%%%%%%%%%%%%%%%%%%%%%%%%%%%%%%%%%%%%%%%%%%
%%%%%%%%%%%%%%%%%%%%%%%%%%%%%%%%%%%%%%%%%%%%%%%%
%%%%%%%%%%%%%%%
\section{Conclusion}

In this work, we have combined nested sampling with both machine-learned and empirical interatomic potentials of lead to explore phase behaviour across a wide range of thermodynamic conditions, giving us insights into the models' properties and highlighting the power of unbiased sampling.

A key outcome of this study is the clear advantage of the neural network-based Pb-EDDP model, trained on DFT data, over empirical potentials for predicting experimentally relevant phase behaviour. Despite their lower computational cost, both the Pb-EAM and Pb-MEAM models fail to reproduce the correct phase behaviour.
While the melting temperature is largely captured well and the low-pressure solid phase is correctly identified as FCC, their reliability rapidly deteriorates with increasing pressure. This is consistent with our past observations in other metallic systems~\cite{ns_iron_2018,NS_lithium,pt_phase_dias_ns}, highlighting that such models lack the flexibility required to remain reliable beyond their original fitting conditions.
In contrast, the EDDP model captures all relevant phases and their relative stability across a wide range of conditions, with results in good agreement with previous work~\cite{salzbrenner2023}. Notably, the sampling does not reveal any apparent fault in the potential energy surface, and the model performs robustly even at high temperatures and pressures beyond those included in the training set.

MLIPs, such as EDDP, are inherently more computationally expensive than empirical models like (M)EAM. Using the same NS parameters, sampling the configuration space of a $30$-atom lead system required more than $100$ times more computational effort with Pb-EDDP compared to Pb-MEAM, consistent with previous studies combining NS with MLIPs.~\cite{ns_carbon, ns_Pt} This increased cost directly limits the accessible system size and sampling resolution. 
Nevertheless, the results presented here demonstrate that even very small systems and coarser sampling parameters are sufficient to recover the key features of the phase diagram with NS, including solid–liquid and solid–solid transitions, as well as the identification of the thermodynamically stable crystal structures, without relying on prior assumptions.
While finite-size effects are present -- most notably in the overestimation of the melting temperature and an enhanced prevalence of stacking variants -- they do not obscure the overall phase behaviour.

Our work also provides useful insights into the performance of the new replica-exchange NS (RENS) scheme, which was found to result in smoother thermodynamic description and sharper phase transitions.
Across all three studied models, a smaller proportion of stacking variants were observed compared to independent NS calculations, but a larger number of configurations were left uncategorised even after the geometry optimisation; this may arise from replica exchanges near the transition region, where liquid-like and solid-like configurations are more readily mixed, although this requires further investigation. RENS was found especially effective at resolving low-temperature solid–solid transitions, where conventional single-pressure NS may lose walkers from one of the competing basins.
Importantly, this improved resolution was achieved at a significantly reduced computational cost (approximately a factor of five lower for empirical models where comparisons could be made), making RENS particularly attractive for MLIP-based studies where computational expense can be a limiting factor for choosing sampling parameters.

More broadly, the combination of NS, and particularly RENS, with modern MLIPs represents a promising route towards direct sampling of the partition function and making thermodynamically relevant predictions to achieve a robust and unbiased view of the configuration landscape. Through these, NS offers a valuable framework for validating MLIPs, providing an unbiased test of their ability to reproduce expected behaviour, while simultaneously generating informative configurations that can be used to iteratively improve the underlying models if needed~\cite{fletcher2025autonomous}. 

%The Pb-EDDP model successfully captures the FCC to HCP transition, in addition to evidencing a further transition to BCC at higher pressure. Given the inability of both EAM models to capture the correct phase behaviour, it is clear that the machine-learned model is worth the additional expense. Opting to minimise the NS configurations in a post-processing step is useful for identifying solid-solid phase transitions, although it can become costly at especially high pressures. Generally, RENS led to smoother phase transitions, and a reduction in the quantity of stacking variants observed.

\begin{acknowledgements}

L.B.P. acknowledges support from the EPSRC through the individual Early Career Fellowship (EP/T000163/1).
Computing facilities were provided by the Scientific Computing Research Technology Platform of the University of Warwick. P.I.C.C. and C.J.P acknowledge funding from the Enterprise Science Fund, Intellectual Ventures. 

\end{acknowledgements}

%\clearpage

%\appendix

\bibliography{NS_basics,LeadRefs,pascals}

\end{document}